# Detecting residues of cosmic events using residual neural network

Hrithika Dodia

Dwarkadas J. Sanghvi College of Engineering, No. U-15, J. V. P. D. Scheme, Bhaktivedanta Swami Road, Vile Parle, Mumbai, Maharashtra, India.

Email address: hrithikadodia19@gmail.com

**ABSTRACT:**

The detection of gravitational waves is considered to be one of the most magnificent discoveries of the century. Due to the high computational cost of matched filtering pipeline, there is a hunt for an alternative powerful system. I present, for the first time, the use of 1D residual neural network for detection of gravitational waves. Residual networks have transformed many fields like image classification, face recognition and object detection with their robust structure. With increase in sensitivity of LIGO detectors we expect many more sources of gravitational waves in the universe to be detected. However, deep learning networks are trained only once. When used for classification task, deep neural networks are trained to predict only a fixed number of classes. Therefore, when a new type of gravitational wave is to be detected, this turns out to be a drawback of deep learning. Shallow neural networks can be used to learn data with simple patterns but fail to give good results with increase in complexity of data. Remodelling the neural network with detection of each new type of GW is highly infeasible. In this letter, I also discuss ways to reduce the time required to adapt to such changes in detection of gravitational waves for deep learning methods. Primarily, I aim to create a custom residual neural network for 1-dimensional time series inputs, which can learn a ton of features from dataset without giving up on increasing the number of classes or increasing the complexity of data. I use the two class of binary coalescence signals (Binary Black Hole Merger and Binary Neutron Star Merger signals) detected by LIGO to check the performance of residual structure on gravitational waves detection.

# INTRODUCTION:

The term "gravitational wave" was coined by Einstein in the year 1916, in his general theory of relativity. After almost 100 years, Laser Interferometer Gravitational-Wave Observatory (LIGO) detected gravitational waves at their Livingston and Hanford detectors. They bagged Nobel Prize for this amazing discovery which opened new doors to our quest of unlocking the yet unknown secrets of the universe. For confirming the detections, LIGO uses matched filtering techniques which are computationally very expensive and also time consuming. Therefore, many researchers have turned towards deep learning for making a better system for this purpose that can make rapid predictions which are necessary especially when GW signals are followed by their electromagnetic counterparts. In 2015, GW150914 [1] was the first confirmed gravitational wave event that originated from merger of two black holes more than a billion light years away. Two years later, in 2017, BNS mergers were discovered as a new source of gravitational waves in the universe through the GW170817 [2] event. Another suspected source of gravitational waves are Core-Collapse Supernova (CCSN) events. With increase in sensitivity of LIGO detectors we can expect many different types of sources of gravitational waves to be discovered leading to various GW signal patterns and more classes of GW required to be classified by systems confirming detections. A neural network is trained only once and after it is trained, it can predict results for any input time series in just seconds. But what happens when we have an entirely new type of GW data coming up like CCSN when our model is trained to predict only BBH and BNS merger signals? In this work, I explore advanced deep learning techniques to tackle this issue. At first, we need a deep neural network which is able to learn various types of gravitational wave signals accurately without facing the need to change structure of the neural network with increase in number of classes or complexity of gravitational waves pattern. This is required because designing and training a deep neural network from scratch is a very time-consuming process and hence it is not convenient to be done at the time of confirming detection. To create such robust network, we need to go deeper by stacking many convolutional layers to increase the complexity of network so that it can sustain large amount of data and a ton of new features. To prevent depth from degrading neural network's performance, I use residual connections.

Deep Residual Learning Framework, popularly known as ResNet, was introduced in 2016 by He *et al.* [3]. What makes it different from traditional deep learning models are skip connections. Skip connections or shortcut connections are those skipping one or more layers and are basic building blocks of ResNet. However, it is made for 2-dimensional data like images. Our data is 1-dimensional time series vectors. I create a 1D Custom Residual CNN for classification of gravitational waves which takes as input raw time series data and provides probabilities that the time series belongs to a specific class as the output. To see how well residual connections can detect gravitational waves, I train my neural network on signals belonging to the two compact binary coalescence classes detected by LIGO till date, with high probability – BBH merger and BNS merger. The second step would be Transfer Learning. There has been a lot of hype around it in the machine learning community due to it's amazing results. Transfer Learning is a technique in which knowledge gained after learning a task is used for learning a relevant task. It can also be applied to gravitational waves. In this paper, I present the details and results of my custom deep residual network and only hypothesize how application of transfer learning can be beneficial as the next step. Transfer learning can be done by initializing the neural network with weights of the pretrained network and then training it (in our case pretrained network is the one trained on

BBH and BNS class). This initialization gives higher accuracy and reduces training time in most of the cases. Hence, it is much better than training the network from scratch. In this way, we can considerably reduce the time required for making existing deep learning systems compatible for detecting the new types of gravitational waves.

**DATASET:**

The pool of LIGO confident detections [1,2,4–10] contain gravitational wave signals originating from merger of two black holes and merger of two neutron stars. With this reason, I confine my dataset labels to three categories:

1. Noise
2. BBH Merger Signals
3. BNS Merger Signals

The dataset is generated using LALSuite library by LIGO. To simulate BBH merger signals, I use IMRPhenomD-type waveform which models inspiral, merger and ringdown components of the binary coalescence. The range of component masses for BBH signals is chosen as $5M_\odot$ to $50M_\odot$ incremented in steps of $1M_\odot$ for training and validation dataset and in steps of $0.5\ M_\odot$ for the test dataset. The training and validation set contains signals with only even SNR in range 2 to 20 whereas test set contains all SNRs (even and odd both) in range 2 to 24. The reason for this is, in real world scenario, there can be detections with masses and SNR different from the training and validation set. To test my network's generalization, the test set parameters are chosen to be completely different from the data used for training the model. The spin of each black hole is considered to be zero. The templates are also replicated over a few different realizations of noise. While simulating BNS merger signals, one more parameter is to be considered – Tidal Deformability. The tidal deformability is the measure of deformation of a body caused by tidal fields which are generated when two massive objects orbit each other. For computing tidal deformability, I use APR equation of state [11]. The tidal deformability of black holes is zero. The BNS merger templates are generated using PhenomPNRT waveform model with component masses between $1M_\odot$ to $2M_\odot$ sampled in steps of $0.02\ M_\odot$ for training and validation set and in steps of $0.01 M_\odot$ for the test set. The SNR range for BNS signals is chosen to be 6 to 36 for training and validation set. The signals are sampled at 4096 Hz and signal duration is chosen to be 10s. The waveforms are scaled according to optimal SNR and injected to noise. The signals are then whitened with Advanced LIGO's power spectral density (PSD) and highpassed at 20 Hz. The training and validation set contain 59671 and 5189 time series respectively. Both of them have approximately equal fractions of each class of data. The figures Figure *1* and Figure *2* show input time series examples.

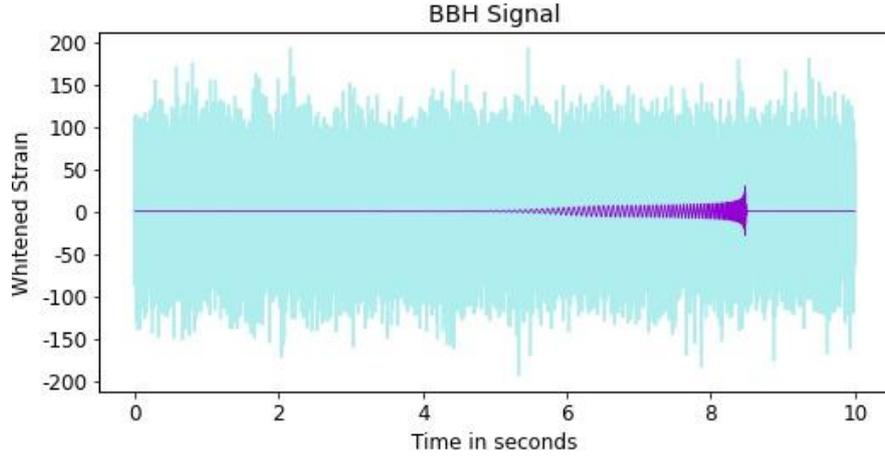

Figure 1: BBH merger signal injected in noise with optimal SNR = 8 is shown in pale turquoise. The component masses of the black holes are $35M_\odot$ and $25M_\odot$ for violet noise-free BBH waveform. The signal in pale turquoise is an example of input to my neural network.

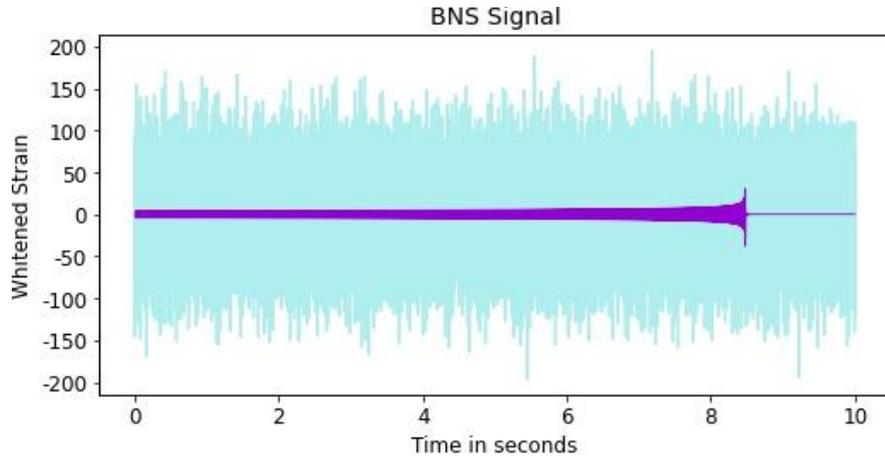

Figure 2: BNS gravitational waves noise-free timeseries with component masses $1.8M_\odot$ and $1.2M_\odot$ is shown in purple. The signal in pale turquoise represents the timeseries obtained after injecting the signal in violet to noise with optimal SNR = 10 and whitening.

**ARCHITECTURE:**

The basic layers in the neural network are convolution, batch normalization, max pooling, average pooling, dropout and dense fully connected layers. I represent the 1-dimensional convolution layer as CONV1D(filters, kernel_size), max pooling layer as MAXPOOL(pool_size) and average pooling layer as AVERAGEPOOL(pool_size). The residual network is made up of identity blocks and convolution blocks. An identity block (see Figure *3*) contains 2 convolution and 2 batch normalization layers. I denote identity block as IDENTITYBLK(f1, f2) where f1 and f2 are the number of filters for first and second convolution layer respectively. The kernel size is 1 for first convolution layer and 3 for second convolution layer of the identity block. The stride is kept 1 for both the convolution layers. In identity block, the shape of output = input shape and the input is simply added to the output of batch normalization after second convolution layer. In convolution block (see Figure *4*), the input is passed through a shortcut component comprising of convolution and batch normalization layer before performing addition at the end of the block. For the convolution layer at start and in shortcut component, the kernel size is 1 and stride is 2. The second convolution layer has stride of 1 and kernel size of 3. I represent convolution block as CONVOLUTIONBLK(f1, f2) where f1 and f2 are filter sizes. Convolution block performs downsampling and hence the output dimensions are not the same as input dimensions.

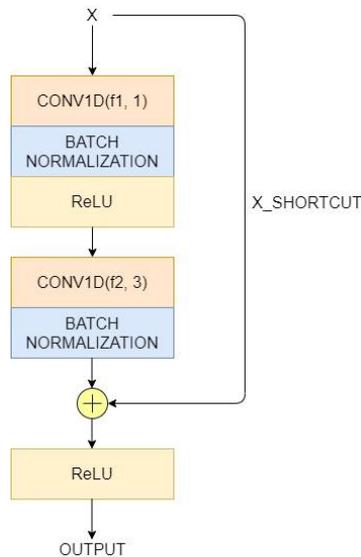
Figure 3: Identity block. The parameters f1 and f2 indicate number of filters and vary for each block.

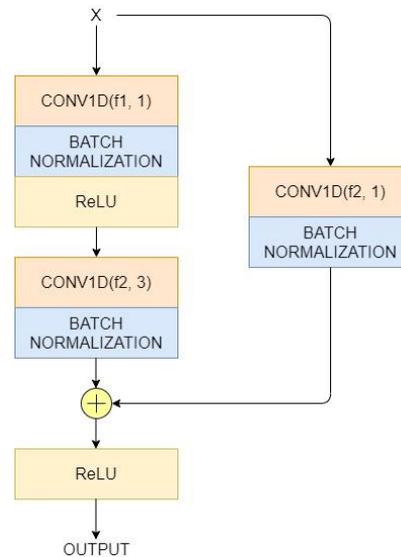
Figure 4: Convolution block. The parameters f1 and f2 indicate number of filters and vary for each block.

In the neural network, the input taken through the input layer is passed to a convolution layer with 64 filters, kernel size of 11 and stride of 2. It is followed by batch normalization and maxpooling layer. After this, convolution and identity blocks are stacked to form a deep neural network. The filter sizes across all the blocks are among 64, 128, 256 and 512. Following the last identity block, there is an average pooling layer with pool size of 8. The output of average pooling layer is flattened and passed through 3 fully connected dense layers separated by 2 dropout layers. The complete structure of my neural network is shown below in Figure 5:

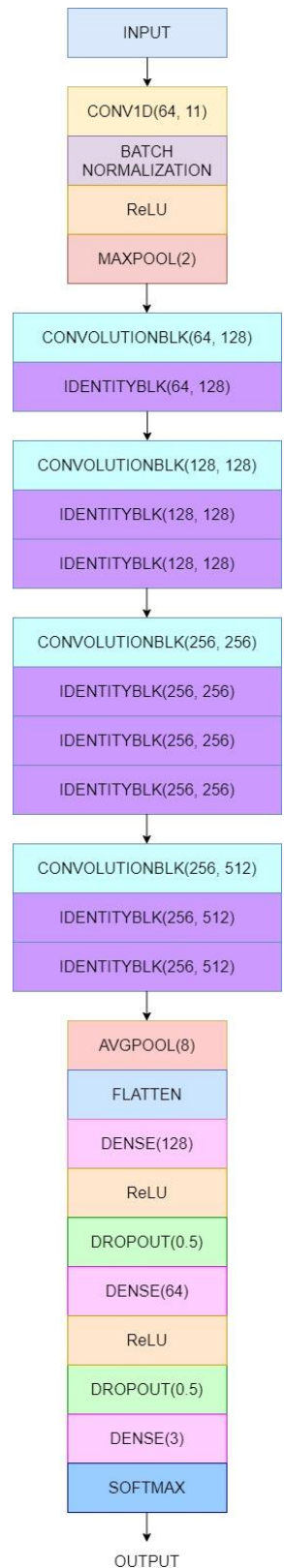

Figure 5: My custom residual neural network architecture for detection of gravitational waves.

This network was created after fine tuning of multiple hyperparameters. The dataset was completely shuffled before training the neural network unlike many papers that implement curriculum learning for training the network. The learning algorithm used was ADAM. I used Keras to build and train this network. The initial learning rate was chosen to be 5e-4. Batch size was 64. The network was trained on a NVIDIA Tesla V100 with 16GB GPU memory on AWS (Amazon Web Services).

**RESULTS:**

For detailed analysis, I assess my neural network's performance on each signal-to-noise ratio (SNR) of BBH and BNS class separately. To do this, my test dataset contains 405 samples of each SNR. Straightforward and effective metrics is used to gauge the performance of my trained classifier. For each SNR, I calculate detection ratio given by, number of signals correctly classified upon total number of signals (which is 405). The detection ratio metrics is very similar to the one used by Gebhard *et al.* [12] and it can also be termed as sensitivity. This will give us clear hint of how well my model is able to trigger the presence of the binary coalescence signals buried in noise. Here, detection ratio = 1 will indicate that all signals are correctly classified for that SNR of the particular signal class. A specific range of masses and SNR is chosen whose combinations are used to create signals with BBH and BNS injections. To test classifier's generalization capability, I include only even SNR signals in the training data. Also, step size for incrementing masses for BBH and BNS signals are different for both training and test dataset. This means that parameters of approximately 50% of the signals in the test set are not even included in the parameter set of the data used for training the network. Apart from this, all the samples in test dataset are not seen by the network before.

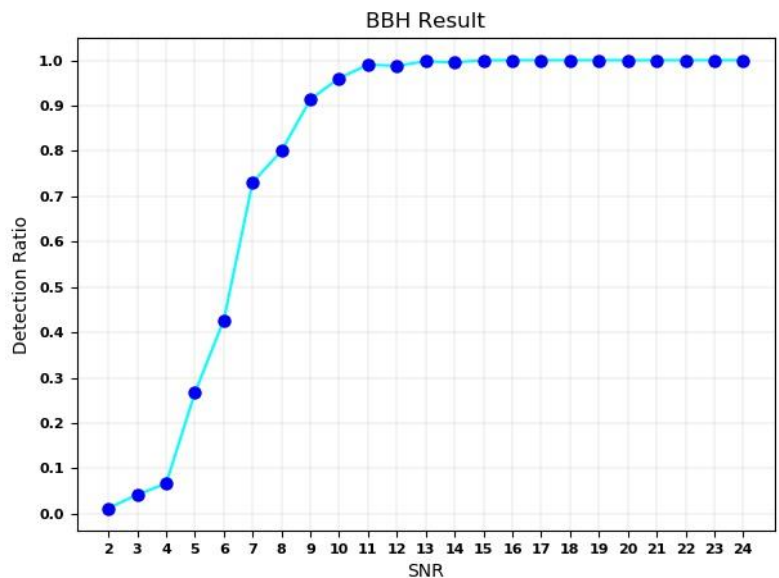

Figure 6: Detection ratio curve for BBH signal test data with optimal SNR ranging from 2 to 24.

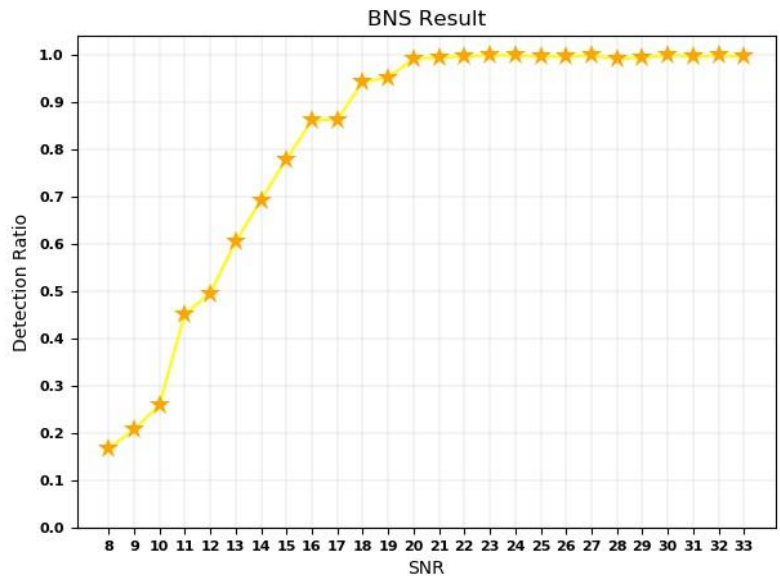

Figure 7: BNS signal test data detection ratio curve with optimal SNR ranging from 8 to 33.

Based on the results, it can be stated that the network is more sensitive in detection of BBH signals than BNS signals. The detection ratio increases steeply with SNR and reaches approximately 100% for BBH signals at SNR = 11. For BNS signals, 100% sensitivity is achieved at SNR = 20. Looking at the constant increase of sensitivity in both the plots Figure 6Figure 6 and Figure 7, we can conclude that the neural network successfully classifies the signals irrespective of whether it is trained on that mass pair and SNR or not. To further increase the sensitivity, we can increase the number of training samples. The network has compressed several gigabytes of data to only 95 MB. For fixed classes, training is required only once and the 95 MB model thus obtained would be used for predicting GW. This gives deep learning an edge over matched filtering, in which every time when there is a suspected detection, data is matched with a large number of templates. This makes matched filtering a very computationally expensive process.

**CONCLUSION:**

At the end, I have successfully demonstrated the use of residual neural network for detection of gravitational waves with raw time series data of BBH merger, BNS merger signals and noise. This very deep neural network can be applied to various other types of GW data such as core-collapse supernova and neutron star-black hole signals. The same architecture can also be used for source parameter estimation from gravitational waves signals by simply removing the final softmax layer. Moreover, this method ensures rapid detection of gravitational waves which is very crucial especially when GW signals are accompanied by electromagnetic radiation like in case of GW170817.

**ACKNOWLEDGEMENT:**

I heartly thank my mom Alpa Dodia, for her constant support and motivation and without whom this research was not possible. In this research, I have made use of software tools provided by Gravitational Waves Open Science Centre (https://www.gw-openscience.org)